# Seeking the Principles of Sustainable Software Engineering

WSSSPE 2014 Workshop

Justin Y. Shi | shi@temple.edu

July 19, 2014

**Abstract**

Like other engineering disciplines, software engineering should also have principles to guide the construction of sustainable computer applications. Tangible properties include a) unlimited scalability, b) maximal reproducibility, and c) optimizable energy efficiency. In practice, we expect a sustainable scientific application should be written once and execute many times on multiple different processing platforms of different scales with optimized performance and energy efficiency.

For more than two decades, explicit parallel programming/processing paradigms only focused on performance. Practices showed that the rigid program-data binding prohibited dynamic runtime resource optimization and fault isolation, making it difficult to reproduce applications in scale.

This paper reports our practice and experiences in search of the first principles of sustainable software engineering for compute and data intensive applications. Specifically, we report our practice and experiences using implicit parallel programming/processing paradigms.

1. **Why Implicit Parallel is Necessary – An Architectural Argument**

Unlike explicit parallel programs, an implicit parallel application allows the runtime layer to decouple the applications from the processing components. This gives the runtime layer a fighting chance to deliver dynamically optimized performance and reliability at the same time. Consequently, not only application reproducibility could be improved, energy efficiency could also be improved and performance scalability limit may be lifted altogether.

As technology developments will inevitably change the future processing, communication and storage methods [3], the explicit parallel paradigms have become counter-productive to scientific research and big data processing applications, especially in light of reproducibility of large scale computer applications [1].

2. **Why Implicit Parallel is Necessary – A Theoretical Argument**

Explicit parallel paradigms rely on the validity of "virtual circuits". While it is commonly believed that a "virtual circuit" is a lossless, error free and order-preserved communication channel between programs, a closer look exposed its inherent vulnerability [2] in extreme scale applications. The culprit is in the single-thread buffer management layer of virtual circuit implementation protocols, such as TCP/IP [26] and SDP [27]. Due to the lack of retransmission discipline in the application logic, any transient software and hardware failure can break a virtual circuit and halt the entire application. Thus larger scale applications suffer progressively worse mean time between failures (MTBF). With this vulnerability, it became very difficult to reproduce a large scale scientific application. The irony is that many modeling deficiencies are only exposed in large scale simulations. Some high profile results had to be recalled. The 2009 Yale University round table discussion [1] raised the question of needing data and code sharing at the same time as a scientific paper is published. However, for small scale applications sharing data and code would be sufficient. Due to the architecture inflection point [3], large scale scientific applications are practically very difficult to reproduce if not impossible.

Figure 1 illustrates the OSI 7-layer communication protocol stack. The data layers are the "Achilles Heel" of all virtual circuit-based applications. In practice, these layers are implemented by a single thread program without error checking and re-transmission. According to the impossibility theory of perfect communication [13 and 22], it is impossible to establish a perfect data communication channel when either hosts can experience transient failure. The lack of retransmission discipline in the application ensures progressively worsening reliability as the infrastructure scales in size. The common believe is flawed.

Also in theory, only an implicit parallel paradigm can isolate the communicating hosts from this vulnerability. Further, only an application level statistic multiplexed data layer similar to the packet layer can promise lossless and error free application level data communications [26].

**Figure 1. Virtual Circuit Vulnerability**

This analysis calls for the design and implementation of a tuple switching network [7, 8, 9 and 14]. Similar to the packet switching network, the tuple switching network can deliver incrementally better application performance and reliability at the same time. Thus, by the principles of mathematical induction, statistic multiplexed computing (SMC) applications can enjoy free scalability without any inflection points [9].

### 3. Why Load Balance is the New Performance

Implicit parallelism has a bad name in delivering poor performance in the past. Explicit parallel programs have demonstrated breath-taking high performance results in large scales. However, the laws of physics dictate that growing system speed and size will increase its instability. Load imbalance is inevitable. Therefore, naive N/P parallelization must have room for further improvements.

In practice, we have found that load imbalance can be demonstrated even in small scale dedicated homogeneous environments [8]. A recent experiment comparing a MPI dense matrix program against a SMC wrapped same MPI program in a small scale 16 processor cluster showed 10% better performance for the granularity tuned (load balanced wrapped MPI program) [10]. While this result may be surprising, the physics is simple. The homogeneous multiprocessors share the same interconnect (in the case of our experiments, we had two Infiniband networks). The speeds of data returns will never be automatically balanced if we assign the work load by the naive equation of N/P. Only implicit data parallel paradigm gives the opportunity for granularity tuning by changing the task sizes until all workers reach an equilibrium in finishing time. Generalization of this principle infers that it may be possible to deliver competitive performance for HPC clouds if granularity tuning becomes possible [20].

## 4. Semantic Correctness

The tuple switching network (TSN) has built-in tuple retransmission protocol similar to the packet switching network. The technique is called "shadow tuple". The shadow tuples record computation state that are designed to counter transient failures. Parallel workers are automatically protected by shadow tuples without checkpoints. Other parallel programs, i.e., masters, may not work correctly. To counter master failures, multiple concurrent master programs can be launched sharing the same global computation states. Checkpoints become unnecessary if the concurrent masters produce acceptable results. A recent study categorized three out of four application stability types that qualify for multiple redundant masters without checkpoints [11]. Figure 2 illustrates these four application types.

|  |  | Output | |
|---|---|---|---|
|  |  | Deterministic | Non-Deterministic |
| Input | Deterministic | Supported. Examples: Linear and nonlinear solvers, FD, Molecular simulation, etc. | Supported. Example: Parallel search, unstable parallel sort, etc. |
| Input | Non-Deterministic | Not Supported. Example: Parallel random search, random date-time based reduction | Supported. Example: Monte Carlo simulations |

**Figure 2. Application Stability Types**

Checkpoint-restart is still necessary for the non-supported application masters. For these applications, the savings are in the parallel workers. They are automatically protected.

## 5. Unlimited Scalability for Data Intensive Applications

When a large scale simulation is helped with observation data points, the result of computation becomes more interesting. The proliferation of mobile and wireless devices with sensors have made "citizen scientist" programs possible. This big data phenomenon challenges the computing architecture models that the future scientific computing will need to support applications that are both compute and data intensive.

Currently, the compute intensive applications have already generated enormous data that all must be stored in a stable storage. Unfortunately, existing storage systems also rely on the virtual circuits in their implementation paradigms. They too suffer the architecture inflection point that for mission critical applications, increasing scale can only gain in performance or reliability, but never both. It is commonly believed that data consistency, service availability and network partition tolerance (CAP Theorem) cannot be expected at the same time [4].

Applying the above mentioned first principles to data intensive applications reveals that the CAP proof [5] also relies on the use of virtual circuit. It is then possible to mitigate CAP limitations using the same statistic multiplexing principles. This calls for the design and implementation of a transaction switching network or TRSN [6 and 12]. Preliminary studies show that CAP limitations can indeed be broken if we apply the first principles correctly in the implicit parallel layer and in the application retransmission discipline [17 and 23]. This means that future compute and data intensive science application could all enjoy scale free development paths while giving technology developers the ultimate freedom to perfect their dream devices. Note that although scalability limit is no longer an inhibiting factor, the economic rule of diminishing of returns still apply. Infrastructure architects are responsible for deployments that makes the most economic sense to a given budget.

## 6. Summary

Larger scale computer applications are inherently more volatile than small scale applications. Static and explicit architectures will not survive in extreme scale deployments, regardless remedial efforts in limited special cases. The emergence of big data phenomenon, large scale science application reproducibility problem and the architectural inflection point claim encourage timely discussions on these important research directional topics.

Automatic harnessing of massive component volatility also makes quantitative parallel application scalability analysis more practical [20]. Data parallelism automatically optimizes deliverable performance without NP-hard algorithms [18]. The SMC framework also brings a new dimension to the traditional Tuple Space paradigm [21] that not only promises to solve the scalability and reproducibility challenges but also give the chance to address the long standing Byzantine failure issues [24 and 23].

As the engineers of the Golden Gate Bridge had anticipated the earth quakes using the first principles, software engineers need also try to discover the first principles for extreme scale application engineering. Emerging large scale distributed systems, such as Hadoop [15], NoSQL, Cassandra, MongoDB and Spark [16], are leading the charge. The proposed SMC paradigm is a continuation of these developments. Since the computing applications will have broad impacts on our social, economic and political futures, the importance of this task cannot be overestimated.

We hope that our preliminary investigative results have provided enough evidence for a necessary paradigm shift.

## 7. Recommended Actionable Items

To help with the paradigm shift, we recommend the following:

   a) Investigate the full potentials of implicit parallel paradigms. There can be more innovative ideas out there that may accomplish even more than discussed above.
   b) Investigate the transformation methods that can help with the transition while allowing communities to continue develop localized custom applications.
   c) Investigate implicit parallel paradigm with heterogeneous multiprocessor with multiple interconnection networks at the same time.
   d) Investigate "multigrid" methods in both modeling and software engineering in order to overcome the inevitable interconnect bottleneck.
   e) Investigate the full potential of performance tuning for HPC clouds.
   f) Investigate the possibility of automatic implicit parallel wrapper generation.